\documentclass[aps, longbibliography, reprint, superscriptaddress]{revtex4-1}

\usepackage{amsmath,amssymb,amsthm,amscd,bm,bbm}
\usepackage{txfonts} 
\usepackage{mathtools}

\usepackage{afterpackage}
\usepackage{hyperref}
\hypersetup{
    colorlinks,
    linkcolor=blue,
    citecolor=blue,
    filecolor=blue,
    urlcolor=blue
}
\usepackage{multirow}
\usepackage{tikz} 
\usetikzlibrary {shapes.multipart}
\usetikzlibrary {shapes.geometric}
\usetikzlibrary {shapes.arrows}
\usetikzlibrary {arrows.meta}
\usetikzlibrary {calc}
\usepackage[normalem]{ulem}
\usepackage{wasysym}

\usepackage{float}
\usepackage{graphicx}
\usepackage[caption=false]{subfig} 
\usepackage{array, booktabs}
	\newcolumntype{P}[1]{>{\centering\arraybackslash}p{#1}} 

\newcommand{\mean}[1]{\langle #1\rangle}

\newcommand{\vekk}[1]{\boldsymbol{#1}}
\newcommand{\f}[2]{\frac{#1}{#2}}

\newcommand{\ra}{\rightarrow}

\DeclareMathOperator*{\Tr}{Tr}

\newcolumntype{C}{>{$}c<{$}}
\AtBeginDocument{
\heavyrulewidth=.08em
\lightrulewidth=.05em
\cmidrulewidth=.03em
\belowrulesep=.65ex
\belowbottomsep=0pt
\aboverulesep=.4ex
\abovetopsep=0pt
\cmidrulesep=\doublerulesep
\cmidrulekern=.5em
\defaultaddspace=.5em
}

\begin{document}

\title{Unsupervised Interpretable Learning of Phases From Many-Qubit Systems}
\date{\today}

\author{Nicolas Sadoune}
\affiliation{Arnold Sommerfeld Center for Theoretical Physics, University of Munich, Theresienstr. 37, 80333 M\"unchen, Germany}
\affiliation{Munich Center for Quantum Science and Technology (MCQST), Schellingstr. 4, 80799 M\"unchen, Germany}

\author{Giuliano Giudici}
\affiliation{Arnold Sommerfeld Center for Theoretical Physics, University of Munich, Theresienstr. 37, 80333 M\"unchen, Germany}
\affiliation{Munich Center for Quantum Science and Technology (MCQST), Schellingstr. 4, 80799 M\"unchen, Germany}

\author{Ke Liu}
\affiliation{Arnold Sommerfeld Center for Theoretical Physics, University of Munich, Theresienstr. 37, 80333 M\"unchen, Germany}
\affiliation{Munich Center for Quantum Science and Technology (MCQST), Schellingstr. 4, 80799 M\"unchen, Germany}

\author{Lode Pollet}
\affiliation{Arnold Sommerfeld Center for Theoretical Physics, University of Munich, Theresienstr. 37, 80333 M\"unchen, Germany}
\affiliation{Munich Center for Quantum Science and Technology (MCQST), Schellingstr. 4, 80799 M\"unchen, Germany}
\affiliation{Wilczek Quantum Center, School of Physics and Astronomy, Shanghai Jiao Tong University, Shanghai 200240, China}

\begin{abstract}
Experimental progress in qubit manufacturing calls for the development of new theoretical tools to analyze quantum data.
We show how an unsupervised machine-learning technique can be used to understand  short-range entangled many-qubit systems using data of local measurements.
The method successfully constructs the phase diagram of a cluster-state model and detects the respective order parameters of its phases, including string order parameters.
For the toric code subject to external magnetic fields, the machine identifies the explicit forms of its two stabilizers.
Prior information of the underlying Hamiltonian or the quantum states is not needed; instead, the machine outputs their characteristic observables. 
Our work opens the door for a first-principles application of hybrid algorithms that aim at strong interpretability without supervision.
\end{abstract}

\maketitle

\section{Introduction}

A shared central theme of quantum computation science and quantum many-body physics is the estimation of a quantum state from a large number of measurements.
Standard quantum state tomography quickly becomes infeasible for this task due to the exponential scaling of the many-body Hilbert space~\cite{Banaszek13, Haah16, Song17}.
Various strategies, such as  matrix product state (MPS) tomography~\cite{Cramer_2010, Lanyon17}, neural network tomography~\cite{Torlai18, Carrasqullia19} and randomized measurements~\cite{Vermersch19, Brydges19, Elben20} have been proposed to improve the efficiency by restricting the target functions to particular information or specific types of quantum states.
A noticeable recent advance in efficiency is the development of shadow tomography~\cite{Aaronson18, Aaronson19} and the classical shadow scheme~\cite{Huang20, Huang21b, Huang21}, which promise to estimate a range of observables accurately from considerably less measurements.
Nevertheless, the resources required to achieve a certain accuracy still strongly depend on the entanglement properties of the target state and the complexity of the observable~\cite{Huang20, Shivam22}.
Therefore, it is highly desirable to develop an algorithm where the defining features of an unknown complex quantum state can be detected and returned as an interpretable output. Those can then be used to make predictions for new measurements and serve as input for more sophisticated tomography processes.

In this work, we show that the tensorial-kernel support vector machine (TK-SVM)~\cite{Greitemann_2019, Liu_2019} can directly analyze positive-operator valued measure (POVM) measurements and identify local entanglement structures in  quantum states, in addition to its proven power of detecting hidden tensorial orders~\cite{Greitemann_2019, Liu_2019} and classical ground-state constraints~\cite{Greitemann19b}.
The \emph{unsupervised} nature and the \emph{strong interpretability} of this machine promise to characterize complex quantum states and construct unknown quantum phase diagrams without knowledge of the states or the Hamiltonian.
We demonstrate the capability of our method by mapping out the phase diagram of a cluster model without supervision, and extracting its analytical string order parameters for arbitrary ranks using experimentally accessible data.
The machine is also applied to a toric code model subject to a magnetic field~\cite{Kitaev03}, where we show that it can distinguish the topologically ordered phase from a trivially disordered phase and identify the stabilizer operators even far away from the zero field limit.\\

\section{Method} \label{sec:Method}

\begin{figure*}[htb]
\centering
\includegraphics[width=0.9\textwidth]{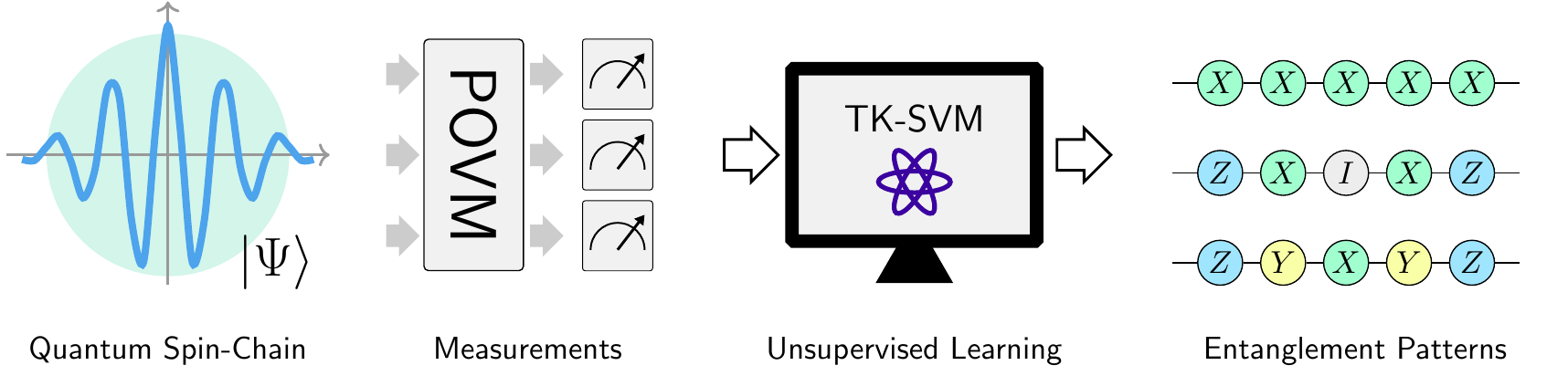}
\caption{Workflow of the tensorial-kernel support vector machine (TK-SVM). The learning stage is unsupervised,
and local correlations patterns as well as local orders in the state can be extracted systematically from the trained model.}
\label{fig:scheme}
\end{figure*}

One of the two uses of our framework is the detection of local observables, that are translational invariant with respect to a unit cell comprised of a number of
adjacent lattice sites. To that end, a Support-Vector-Machine (SVM) is trained to classify the input samples against a set of fictitious featureless
samples. Once the training is completed, the local observables characterising the input are retrieved from the learning-model
internals. In this context, {\it interpretability} means that the model internals have a
direct physical meaning. This is in sharp contrast to neural-network based methods which often lack interpretability.
The second use of TK-SVM consists in exploring completely unknown phase diagrams. For this purposes multi-classification
of a predefined grid of phase points is carried out, followed by partitioning of a graph that is built from learning-model 
parameters as weighted edges and phase points as vertices.\\
Although standard SVM is a supervised learning method, TK-SVM can be seen as unsupervised, since
the labelling of training samples used in TK-SVM, carries no information at all about the physical quantities
(order parameters) that are being learned.
The TK-SVM working scheme is depicted in Fig.~\ref{fig:scheme}.\\

{\it Data and Classical Map:}
From a many-qubit system, repetitive POVM measurements are taken to serve as input to TK-SVM.
Informational completeness (IC) of measurements is crucial to our approach, which is why we
require POVM outcomes as input data (details in Appendix~\ref{sec:POVM}).
The choice of the IC-POVM does not affect the algorithm or the results. Therefore we consider the experiment-friendly and intuitive
Pauli-$6$ POVM with $6$ possible outcomes, defined by a set of positive semi-definite Hermitian matrices
$M^{\uparrow \alpha} = \frac{1}{3} \vert \uparrow_\alpha \rangle \langle \uparrow_\alpha \vert$ and
$M^{\downarrow \alpha} = \frac{1}{3} \vert \downarrow_\alpha \rangle \langle \downarrow_\alpha \vert$
with $\alpha\in\{x,y,z\}$ and $\{\vert \uparrow_\alpha \rangle, \vert \downarrow_\alpha \rangle\}$ representing the eigenbases of the Pauli operators $X$, $Y$ and $Z$.
Every possible outcome is assigned to a $3$-component classical vector $\mathbf{S}=(S^x,S^y,S^z)$, 
{\it e.g.}, $M^{\uparrow x} \mapsto (+1,0,0), M^{\downarrow x} \mapsto (-1,0,0), M^{\uparrow y} \mapsto (0,+1,0)$.
Measuring each qubit of the system in a random Pauli basis and mapping the outcomes yields one classical snapshot of a quantum state.
This is repeated a number $N_s$ of times to obtain a set of samples used as input data.\\

{\it Feature Vector Construction:} The SVM at core of our framework doesn't directly operate on classical spin configurations,
but on a set of feature vectors constructed from these configurations instead. Two hyper-parameters, {\it cluster} 
-- a collection of adjacent lattice sites -- and {\it rank} generally determine the way that a feature vector is constructed.
In this section, only the chain lattice is considered, as the generalization to other
lattice geometries is straight-forward.
For the chain lattice, there is only one possible cluster shape: the string. Therefore the relevant hyper-parameters
become string length $n$ and rank $r$.
The optimal choice of $n$ and $r$ depends on the precise nature of the local orders 
present. Nematic order for instance, can by definition never be captured at rank $r=1$. In practice finding the optimal 
choice of hyper-parameters requires multiple TK-SVM runs on a trial-and-error basis. Often it is also necessary to combine
results from several different choices of hyper-parameters to obtain a complete physical picture. The latter is the case
for the cluster model.\\
For a fixed choice of hyper-parameters, a classical spin configuration of $L$ sites is partitioned into non-overlapping clusters
$\{S^c_i\} \longmapsto \{S_{J,\alpha}^c\}$
where $i\in\{1,...,L\}$ labels the lattice sites, $c\in\{x,y,z\}$ the spin-component, $J\in\{1,...,L/n\}$ labels the clusters 
and $\alpha\in \{1,...,n\}$ the site within one cluster. Note that for the partitioning to be meaningful, $L/n$ must be integer. 
For every cluster $J$, a feature vector of rank $r$ is then constructed by taking all possible 
products of different $S^c_{J,\alpha}$ with exactly $r$ factors.
This yields a set of feature vectors $\{\vekk{\phi}_J\}$ upon which, in a last step, the cluster average 
is taken to produce an averaged feature vector $\vekk{\phi}$
\begin{align}
\vekk{\phi}_J &= \{S^{c_1}_{J,\alpha_1}\times S^{c_2}_{J,\alpha_2} \times ... \times S^{c_r}_{J,\alpha_r}\}  \label{eq:phiJ} \\
\vekk{\phi} &= \frac{n}{L} \sum_J \vekk{\phi}_J. \label{eq:phiAvg}
\end{align}
To exclude trivial products in Eq.~\eqref{eq:phiJ}, any two site indices,
e.g. $\alpha_1$ and $\alpha_2$ must be different. This condition forbids $r$ to be greater than $n$.
Additionally, redundancies are removed from Eq.~\eqref{eq:phiJ} by imposing an ordering $\alpha_1 < \alpha_2 < ... < \alpha_r$
on the site indices. This simplification is allowed since Pauli-operators on different sites commute.
The dimension of the resulting feature vector is independent of the system size $L$ and given by
$\dim (\vekk{\phi}) = \binom{n}{r}\times 3^r$.\\
The procedure above describes how to construct one feature vector from one sample. It is however desirable to create a single
feature vector from several samples, by extending the average in Eq.~\eqref{eq:phiAvg}, to enhance accuracy.
Especially with increasing rank, the cluster average should be taken
over increasingly many samples. This ensures that the error of the components of $\vekk{\phi}$, which can be seen as estimators of
$r$-point correlators, stays arbitrarily small. For more details on the feature vector construction see Appendix~\ref{sec:Clustering}.\\

\subsection{Learning Local Orders (Binary classification)}
When used to learn local orders, the input samples constitute the first class of the training set,
whereas the second class of the training set is constructed internally. With the aim
to learn physical quantities only from the input samples, the artificial second class is chosen to lack any sort of information.
Physically the artificial class might be interpreted as samples from a system at infinite temperature. To emulate such a system,
POVM outcomes are generated randomly according to a uniform distribution. After the artificial samples are generated,
feature vectors are constructed as described in the previous section. This results in two labelled sets of feature 
vectors that can now be fed to the SVM.\\
In our approach, besides quantifying the learning success (Appendix~\ref{sec:Accuracy}),
the decision function is never used to predict unknown samples. Instead the focus lies solely on
the information gained from the input samples, which is encoded in the learning-model internals after training.
These model internals are extracted from the decision function and interpreted as physical observables, characterising the input samples.\\
The quadratic-kernel SVM decision function predicting the class of an unknown feature vector $\vekk{\phi} = \{\phi_\mu\}$ reads
\begin{align}
d(\vekk{\phi}) 
& = \sum_s \lambda_s y_s \big( \vekk{\phi}^{(s)}\cdot\vekk{\phi} \big)^2 - b \label{eq:decisionFunc_1} \\
& = \sum_{\mu\nu} \bigg(\sum_s \lambda_s y_s \phi^{(s)}_\mu \phi^{(s)}_\nu\bigg)  \phi_\mu \phi_\nu - b \label{eq:decisionFunc_2} \\
& = \sum_{\mu\nu} C_{\mu\nu} \phi_\mu \phi_\nu - b \label{eq:decisionFunc_3},
\end{align}
where $s$ runs over the combined training set of feature vectors from both classes, $y_s=\pm 1$ represents the class 
label of each feature vector, and $\lambda_s$ as well as the bias $b$ are the SVM optimisation parameters.
The resulting coefficient matrix $\{C_{\mu\nu}\}$ depends on the training data and optimization parameters only,
and exposes all local orders present in the phase of interest (that the input samples where taken from).
From each of it's non-vanishing components,
an analytical expression interpreted as order parameter or part of an order parameter can be recovered.
Typically a small subset of columns or rows of $C_{\mu\nu}$ is sufficient for inferring an order parameter, and
constructing the entire matrix is not necessary (see Appendix~\ref{sec:Clustering}).
This allows us to examine large clusters and high ranks, which is particularly useful for detecting non-linear and extended orders.\\

\subsection{Learning a Phase Diagram (Multi-classification)}
In general a model's phase diagram might be completely unknown. In this case it is simpler to first get an idea
of the phase diagram's topology by roughly determining phase boundaries,
instead of extracting order parameters of randomly selected phase points.
Using labelled samples from two separate phase points $A$ and $B$ as input (without constructing the artificial class), TK-SVM allows
to decide whether they belong to the same phase, without explicitly extracting local orders from the decision function.
Note that the samples are labeled without knowing if they are truly physically different or not, meaning that this application of
TK-SVM is still unsupervised. The decision whether two classes belong to the same phase is based on a criterion derived
in~\cite{Jonas_thesis},

\begin{equation} \label{eq:BiasCriterion}
|b_{A,B}|
\begin{dcases*}
\gg 1  &\text{if } A, B \text{ in the same phase}\\
\lesssim 1  &\text{if } A, B \text{ in different phases}
\end{dcases*}
\end{equation}

which relies on the bias $b$ of the decision function Eq.~\eqref{eq:decisionFunc_1} alone.\\
To find the approximate topology of an unknown phase diagram, we first fix a grid of $M$ phase points.
Then, for all possible $M(M-1)/2$ different pairs of phase points TK-SVM runs a binary classification task, returning a set of biases.
Succeeding the classification stage, a graph is constructed using phase points as vertices and the set of absolute biases as weighted edges.
Since all points of the same phase are strongly connected (large bias modulus), and points of different phases are
weakly connected, partitioning the resulting graph directly yields the topology of the phase diagram. There are many standard methods to
partition a graph; the one used in this work is the spectral graph partitioning method introduced by Fiedler~\cite{Fiedler73,Fiedler75}
(summarized in Appendix~\ref{sec:GraphPartitioning}).
In practice, this protocol for unsupervised learning of a phase diagram turns out to perform better than the kernel principal component analysis
(PCA)~\cite{Jonas_thesis} and persistent homology~\cite{Olsthoorn20} methods.\\

\section{Applications}

\subsection{Cluster Model}
To demonstrate the abilities of TK-SVM, we consider a spin-$1/2$ Hamiltonian including a paramagnetic, a symmetry-broken, and a symmetry-protected topological (SPT) phase,
\begin{equation}
H= - \sum_{i=2}^{L-1} Z_{i-1}X_{i}Z_{i+1} - h_1 \sum_{i=1}^L X_i - h_2 \sum_{i=1}^{L-1} X_i X_{i+1},
\label{eq:hamiltonian}
\end{equation}
where the sums run over the lattice sites of a chain with open boundary conditions, and the external field $h_1 \ge 0$.
Distinct phases of the Hamiltonian can be understood from the following limits:
For $h_1 \rightarrow + \infty$ the system is a paramagnet with all spins pointing along $\hat{x}$, while for $h_2 \rightarrow - \infty$ it is an Ising antiferromagnet along $\hat{x}$.
The ground state at $h_1=h_2=0$ is known as the cluster state protected by a $\mathbb{Z}_2\times\mathbb{Z}_2$ symmetry and plays a crucial role in measurement-based quantum computations~\cite{Raussendorf_2001, Nielsen04, Verstraete_2004, Nielsen06}.

\begin{figure}[t]
	\centering
	\includegraphics[scale=0.5]{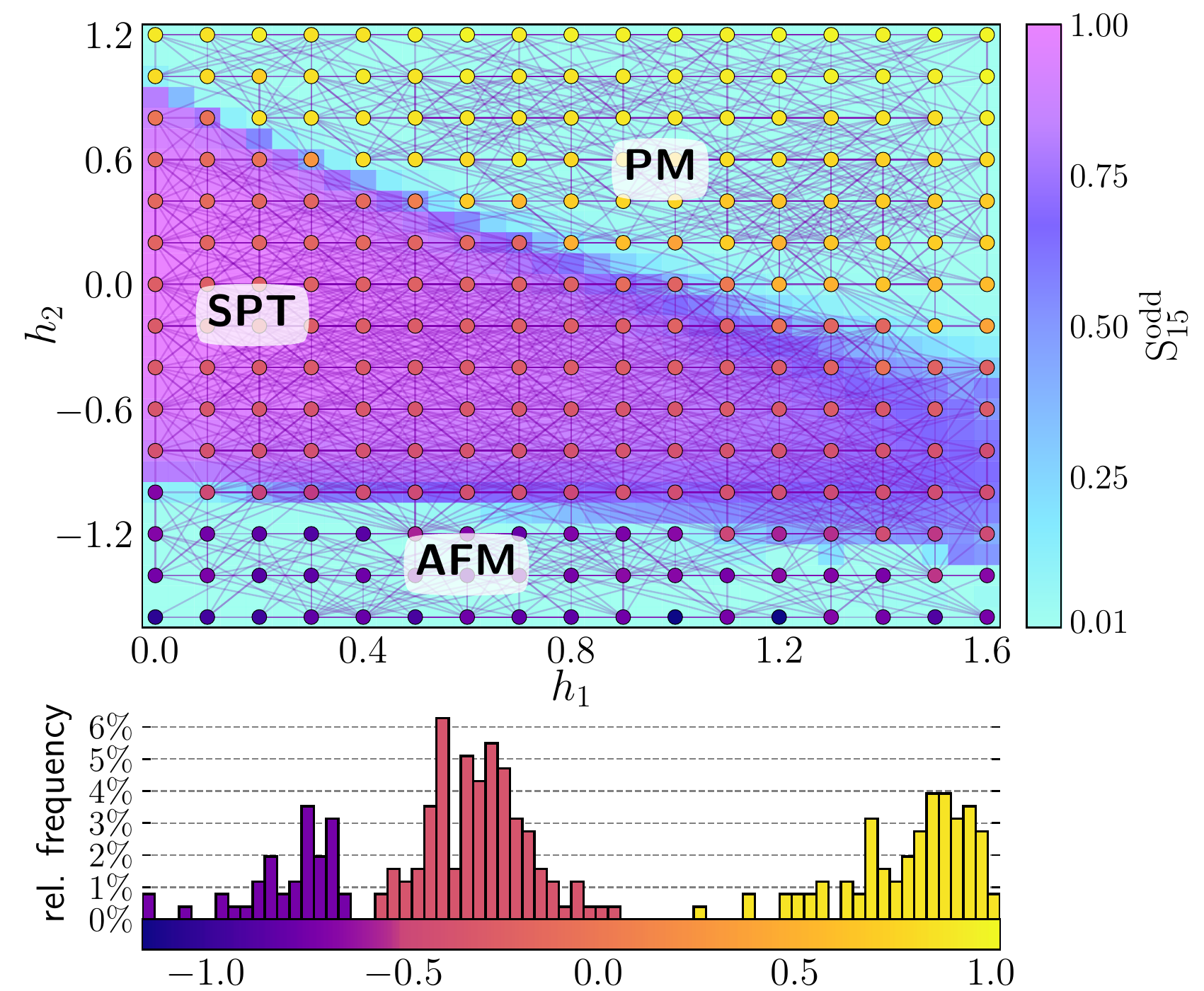}
	\caption{Upper panel: Partitioned graph corresponding to the cluster model phase diagram, learned at rank $r=3$ and cluster size $n=3$.
			 The graph consists of a uniform $15\times 17$ grid of phase points (filled circles) as vertices, and the set of
			 normalized bias parameters, obtained from all mutual classification tasks between the phase points, as edges.
		     Partitioning the graph leads to three components (subgraphs), which are identified as the paramagnetic (PM), SPT ordered,
		     and anti-ferromagnetic (AFM) phases. Every phase point is colored according to the appropriate Fiedler vector entry, which
		     indicates it's phase correspondence.
			 The background diplays the expectation value of the string order parameter
		     $S^{\rm odd}_{n} = \langle Z_1 \big(\prod_{k=1}^{\frac{n-1}{2}} X_{2k} \big) Z_n \rangle$ with length $n=15$,
		     which is finite in the SPT phase but vanishes elsewhere.
		     Lower panel: A histogram of the Fiedler vector entries (for details see Appendix~\ref{sec:GraphPartitioning}),
		     whose values are color-coded (the horizontal axis). The three separated regions correspond to the three phases.}
	\label{fig:phaseDiagram}
\end{figure}

The data used in this work is generated from density matrix renormalization group (DMRG) simulations~\cite{Schollwoeck11},
nevertheless we expect the same analysis to apply for POVM measurements of experimental origin.
A total of $255$ datasets, each containing $N_s = 5000$ POVM samples,
are collected uniformly in a region spanning the whole phase diagram, with system size $L=72$.
Since a Hamiltonian can generally host distinct types of phases and a single phase may also develop more than one order,
two strategies may be used to construct the phase diagram.
The first one is to carry out a multi-classification over the entire dataset with different ranks and clusters, each covering several types of orders.
The full phase diagram is then obtained by combining the results of the subsequent graph partitioning.  
Alternatively, one can work iteratively by restricting the multi-classification to datasets belonging to the same component (subgraph) of a preceding classification analysis, and repeat until a converged phase diagram topology is reached.
Both strategies are more efficient than finding a universal rank and cluster sensitive to all orders, and they can be combined in practice.
In the current problem, we find that rank $r=3$ and cluster size $n=3$ reproduce the expected phase diagram, as shown in Fig.~\ref{fig:phaseDiagram}. 
Increasing $r$ and $n$ does not reveal a finer partitioning, hence no other phases are found.

\begingroup
\renewcommand*{\arraystretch}{1.2}
\begin{table}[t]
\centering
    \begin{tabular}{ c | c | c | c }
    \toprule
    \hline
    ${}\ r,n\ $ & $\dim(\vekk{\phi})$ & $N_f$ & Example \\
    \hline
	$3,6$ & $540$   & $4$ & $B_4=Z_3 X_4 Z_5$ \\
	$4,5$ & $405$  & $3$ & $B_3 B_4=Z_2 Y_3 Y_4 Z_5$\\
	$5,7$ & $5103$ & $8$ & $B_2 B_4 B_6=Z_1 X_2 X_4 X_6 Z_7$\\
	$6,9$ & $61236$ & $30$ & $B_2 B_4 B_5 B_7=Z_1 X_2 Y_4 Y_5 X_7 Z_8$\\
	$7,9$ & $78732$ & $35$ & $B_2 B_3 B_5 B_6 B_8=Z_1 Y_2 Y_3 Y_5 Y_6 X_8 Z_9$\\
	$8,9$ & $59049$ & $22$ & $B_2 B_3 B_4 B_6 B_7 B_8 = Z_1 Y_2 X_3 Y_4 Y_6 X_7 Y_8 Z_9$\\
	$9,9$ & $19683$ & $6$ & $\prod_{k=2}^8 B_k = - Z_1 Y_2 X_3 X_4 X_5 X_6 X_7 Y_8 Z_9$\\
	\hline
	\bottomrule
  \end{tabular}
\caption{Excerpt of non-trivial features learned at different ranks and cluster sizes $r, n$. The dimension of the feature space is
		 $\dim (\vekk{\phi}) = \binom{n}{r}\times 3^r$. $N_f$ denotes the number of non-vanishing features at fixed $r$ and $n$.}
\label{table:BkOps}
\end{table}
\endgroup

\begin{figure}[t]
	\centering
	\includegraphics[scale=0.5]{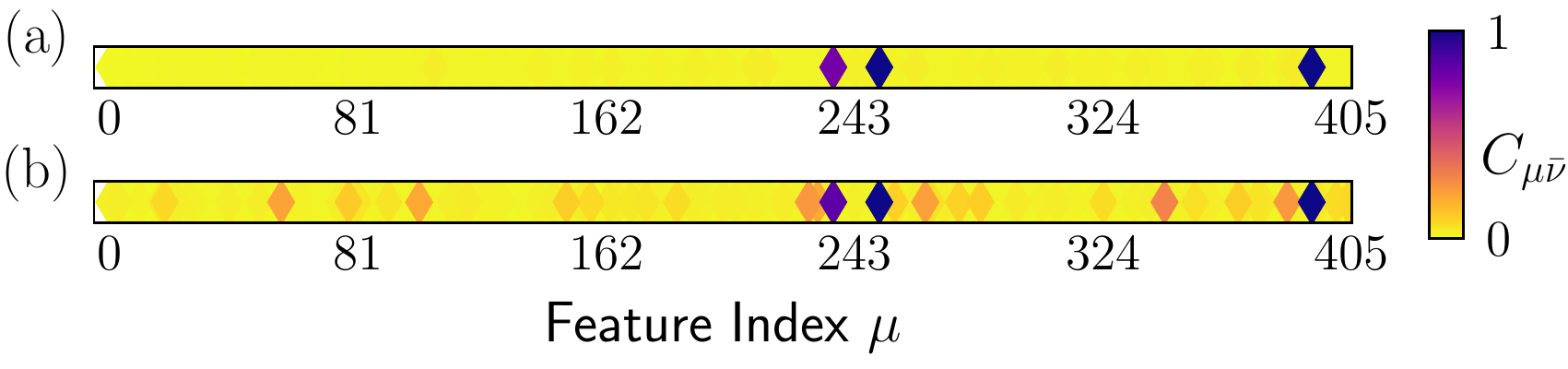}
	\caption{Non-trivial features in the SPT phase learned at $r=4, n=5$.
	The figures show a non-vanishing column of the associated $C_{\mu\nu}$ matrix, with $\bar{\nu}$ chosen
	such that $\phi_{\bar{\nu}} = Z_2 Y_3 Y_4 Z_5$ is a prominent feature.
			 (a) The pure cluster state in the limit $h_1=h_2=0$. Non-vanishing features (diamonds)
			  can be systematically retrieved: the three apparent features  correspond to entangled patterns
			  $B_2 B_4, B_2 B_3$, and $B_3 B_4$ (c.f. Eqs.~\eqref{eq:r4feature_1} and~\eqref{eq:r4feature_2}). 
			 (b) All phase points belonging to the SPT phase according to the graph partition are pooled and treated as one dataset.
			 	The features from the pure limit remain dominant, as they reflect the local correlation structure of the whole phase.}
	\label{fig:FeatureVectors}
\end{figure}

After obtaining the phase diagram, we wish to investigate the order parameters characterizing each phase.
We focus our discussion on interpreting the SPT phase, since the paramagnetic and antiferromagnetic phases only lead to very simple
features $\mean{\sum_{i=1}^L X_i} = 1$ and $\mean{\sum_{i=1}^{L/2} X_{2i-1} - X_{2i}} = 1$.
We start by discussing the pure cluster model in the limit $h_1=h_2=0$ and analyze ranks and clusters up to $r, n = 9$.
The first non-trivial feature emerges at $n=r=3$ with a structure
\begin{equation}
B_k :=  Z_{k-1} X_k Z_{k+1},
\end{equation}
where $k=2,\dots, n-1$ labels qubits within the cluster.
Increasing the rank to $r=4$ and cluster size to $n\geq 5$, two additional structures are detected (see Fig.~\ref{fig:FeatureVectors}),
\begin{alignat}{2}
&B_kB_{k+1}=Z_{k-1} Y_{k} Y_{k+1} Z_{k+2} \quad &&k\in\{2,...,n-2\}, \label{eq:r4feature_1} \\
&B_kB_{k+2}=Z_{k-1} X_{k} X_{k+2} Z_{k+3} \quad &&k\in\{2,...,n-3\}. \label{eq:r4feature_2}
\end{alignat}
In Table~\ref{table:BkOps} we summarize the features learned for ranks and cluster sizes $r,n \in\{3,...,9\}$.
Extrapolating $n, \, r \ra \infty$, they reproduce the commonly
used ``dense'' and ``odd'' string order parameters~\cite{Verstraete_2004, Smith_2022}
\begin{align} \label{eq:denseString}
S^{\rm dense}_{n\ra\infty}
&= \langle \prod_{k=2}^{n-1} B_k \rangle 
= (-1)^n \langle Z_1 Y_2 \big(\prod_{k=3}^{n-2} X_k \big) Y_{n-1} Z_n\rangle \\
\label{eq:oddString}
S^{\rm odd}_{n\ra\infty}
&= \langle \prod_{k=1}^{\frac{n-1}{2}} B_{2k} \rangle 
= \langle Z_1 \big(\prod_{k=1}^{\frac{n-1}{2}} X_{2k} \big) Z_n\rangle.
\end{align}
Although the high-rank features can be generated by the building block $B_k$, it is remarkable that a machine automatically picks up all variants of the string correlations.

To find the order parameter for arbitrary parameter values of $h_1,h_2$, all data sets belonging to the same phase according to the
machine-learned phase diagram are pooled.
For the SPT phase the resulting pooled data set contains $96 \times 5000$ samples.
Classifying the pooled data set against random samples reveals that the same operators as in Table~\ref{table:BkOps}
remain dominant throughout the phase.
Secondary features caused by finite $h_1, h_2$ are significantly weaker, see Fig.~\ref{fig:FeatureVectors}.
We thus identified characteristic local entanglement structures in a quantum phase, without relying on particular known limits
nor any other information about the Hamiltonian.
This marks a crucial difference of our machine compared to popular strategies in
neural-network based algorithms~\cite{Cong19, Bohrdt19, Kottmann20}, which make use of special known limits during the training stage.

\subsection{Toric code subject to magnetic fields}
Characterizing long-range entangled topological phases, \emph{e.g.}, by identifying their anyonic statistics~\cite{Kitaev03, Kitaev06} or topological classes~\cite{Lan18, Lan19}, in a purely data-driven manner, arguably remains beyond the capabilities of existing machine learning algorithms~\cite{note}.
Nevertheless, information about local correlation patterns can bring valuable insight into the state and guide further tomography processes specialized for topological features.
To examine what information can be learned from local POVM data, we consider a toric code model subject to magnetic fields $h_x, h_z > 0$~\cite{Kitaev03},
\begin{equation}
H = - \sum_v \prod_{i \in v} Z_i - \sum_p \prod_{i \in p} X_i - h_x \sum_i X_i - h_z \sum_i Z_i.
\label{eq:toric_ham}
\end{equation}
Qubits are located on the bonds of a square lattice, and $v$ and $p$ denote vertices and plaquettes of the lattice.
This model is topologically ordered for $h_x=h_z \lesssim 0.34$ if $h_x = h_z$ and $h_x \, (h_z) \lesssim 0.33$ if $h_z \,(h_x) = 0$~\cite{wu2012}.
Training data are produced by exactly diagonalizing the Hamiltonian Eq.~\eqref{eq:toric_ham} on a periodic system with $18$ qubits.

\begin{figure}[t]
  \centering
  \includegraphics[scale=0.5]{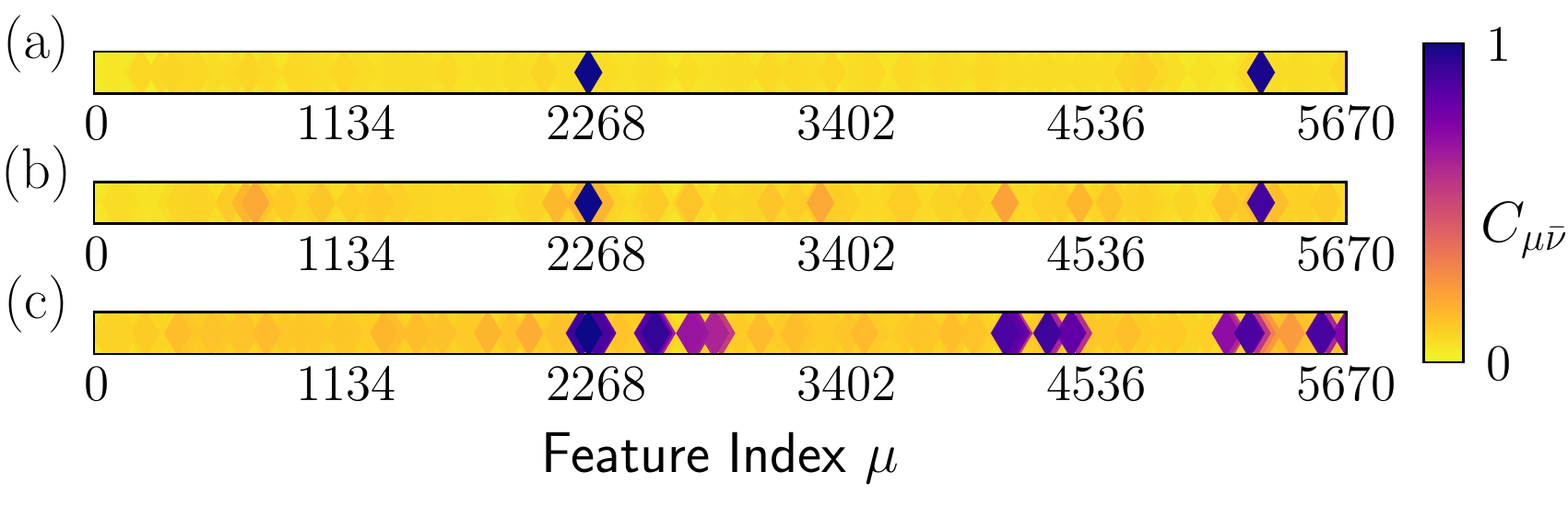}
  \caption{Features detected by TK-SVM for the toric code model using a 
  			$2\times 2$-unit cell ($8$ sites) cluster at rank-$4$. The prominent coefficient matrix column $\bar{\nu}$ is given by 
			$\phi_{\bar{\nu}}=A_{v}$ with $v$ the only vertex contained in an 8-site cluster.
		   (a) In the zero field limit $h_x, h_z \rightarrow 0$, two features (blue diamonds) reflecting the vertex 
		   and plaquette operators $A_v$ and $B_p$ are captured.
		   (b) The same features remain prominent even at sizeable magnetic fields $h_x = h_z = 0.3$ close to the phase boundary.
		   (c) In the non-topological phase at $(h_x,h_z) = (0,1.5)$, multiple trivial features appear due to the strong polarization.}
  \label{fig:toricFeatureVectors}
\end{figure}

We first investigate the pure toric code limit $h_x = h_z \rightarrow 0$.
$N_s = 1000$ Pauli-$6$ POVM snapshots are collected and discriminated against random samples.
We find that using a cluster of $2\times 2$ unit cells ($8$ qubits, as in Appendix~\ref{sec:Clustering}) at rank-$4$,
the machine captures two features whose interpretation yields  the toric code stabilizers $A_v = \prod_{i \in v} Z_i$ and $B_p = \prod_{i \in p} X_i$, as shown in Fig.~\ref{fig:toricFeatureVectors}.
The same procedure is carried out at sizeable fields $(h_x, h_z) = (0.3, 0.3)$, where we identify the same prime features in contrast to the trivial paramagnetic phase at $(h_x, h_z) = (0, 1.5)$ as can be seen in Fig.~\ref{fig:toricFeatureVectors}.
The machine can successfully detect the correct stabilizers from local measurements sampled far away from the pure toric code limit.   
Although stabilizers do not directly characterize the topological phase, their explicit forms indicate the underlying gauge structure and can inspire specialized feature mapping and kernel designs.
For instance, one may generate training sets by sampling closed loops~\cite{Zhang17b, Sehayek22} or define the kernel of a gauge symmetry~\cite{Rodriguez19, Scheurer20}, which are strategies employed in previous studies of machine-learning intrinsic topological orders.

\section{Summary}
Machine learning techniques exhibit growing abilities of analyzing complex classical and quantum data. In this work, we demonstrated the potential of TK-SVM as a \emph{first-principle} method to detect entanglement structures in many-body qubit systems.
We constructed the phase diagram of the cluster model, without supervision, from experimentally accessible POVM snapshots. 
Furthermore we extracted the respective order parameters of the phases systematically, and, in particular,
discussed the local entanglement patterns and string order parameters of its SPT phase.
Finally, we examined a intrinsic topological phase subject to magnetic fields and showed the machine's ability to detect the explicit stabilizers with data far away from the pure limit.
Our results pave the way to investigate membrane-like order parameters in higher dimensional SPT phases~\cite{Chen13} and
analyze local measurements of general topological models such as lattice gauge theories~\cite{Kogut79} and fracton models~\cite{Vijay16}.

\begin{acknowledgments}
We acknowledge support from FP7/ERC Consolidator Grant QSIMCORR, No. 771891, and the Deutsche Forschungsgemeinschaft 
(DFG, German Research Foundation) under Germany's Excellence Strategy -- EXC-2111 -- 390814868. The research is part of the
Munich Quantum Valley, which is supported by the Bavarian state government with funds from the Hightech Agenda Bayern Plus.
Our simulations make use of the $\nu$-SVM formulation~\cite{Scholkopf00}, the LIBSVM library~\cite{Chang01, Chang11} and the iTensor
library~\cite{itensor}.
\end{acknowledgments}

\appendix

\section{IC-POVM and Classical Map} \label{sec:POVM}

Over a Hilbert space of dimension $d$, a general POVM is defined as a set of positive-semidefinite operators $\vekk{M}=\{M^i\}$,
called `outcomes', which sum to identity $\sum_i M_i=I$.
To be informationally complete (IC), the outcomes of a POVM must span the whole space of self-adjoint operators acting on
the Hilbert space. This can only be the fulfilled if the POVM has at least $d^2$ outcomes; if it has exactly $d^2$ outcomes
it is called minimal IC-POVM.
Moreover, a POVM is symmetric if the overlap between any two different outcomes is constant $\Tr(M^i M^j) = const$.
Informational completeness ensures that the density matrix can be uniquely determined from the probabilities $\Tr(M^i\rho)$
and is therefore a strict necessity for our algorithm. On the other hand, POVM symmetry is completely optional for TK-SVM.
Given an IC-POVM, every outcome is mapped onto a classical vector $\vekk{S}\in \mathbb{R}^{d^2-1}$ serving as representation 
of a single site quantum state, the density matrix of which is determined by $d^2-1$ real parameters.
Specifically for the spin $1/2$ ($d=2$) case, the classical one-to-one map $M^i \mapsto \vekk{S}^i\in \mathsf{S}^2$
(onto the Bloch sphere) reads
\begin{equation}
    M^i=\frac{d}{|\vekk{M}|}\rho(\vekk{S}^i)=\frac{1}{|\vekk{M}|} (I + \vekk{S}^i \cdot \vekk{\sigma}),
\end{equation}
where $\vekk{\sigma}$ denotes the Pauli-vector.
A straight-forward choice for an IC-POVM is the Pauli-6 POVM, defined by the set of Bloch vectors
$\{(\pm1, 0, 0)^T, (0, \pm1, 0)^T, (0, 0, \pm1)^T\}$. Although it is not minimal, it has the advantage of being easy to
implement in experiments by simply measuring each site in a randomly chosen Pauli basis.\\
Alternatively, one might use the Tetra-POVM, which is minimal and symmetric (SIC-POVM), defined by the set
$\{(0, 0, 1)^T,$ $(2 {\sqrt{2}}/{3}, 0, -{1}/{3})^T,$ $(- {\sqrt{2}}/{3}, -\sqrt{2/ 3}, -{1}/{3})^T,$
$(- {\sqrt{2}}/{3},  \sqrt{2/ 3}, -{1}/{3})^T\}$, spanning a tetrahedron.
Given that a certain POVM is IC, if all of its Bloch vectors can be expressed as linear combinations of the vectors 
defining a second POVM, then this second POVM is IC as well.
Using either the Tetra- or the Pauli-6 POVM yields fully consistent results in our application of TK-SVM.

\section{Clustering and Feature Detection} \label{sec:Clustering}

\begin{figure*}[htb]
\includegraphics[scale=0.9]{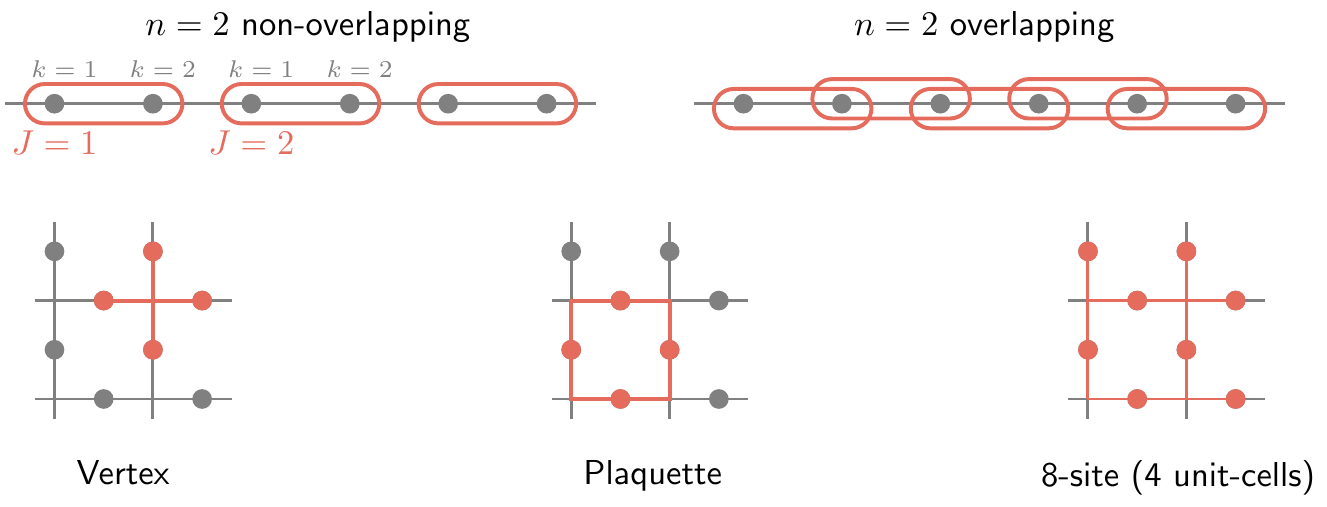}
\caption{Different examples of clusters for the chain lattice (top) and square link lattice (bottom).
		For an unknown quantum state, natural choices for a cluster include multiple lattice cells and the nearest neighbors of a site.
The clusters may or may not be overlapping. Non-overlapping clusters are appropriate for detecting structures with translational symmetry breaking, while overlapping clusters are more suitable for probing structures with translational invariance.
		In case of non-overlapping clusters of the chain, the lattice size $L$ must be divisible by $n$,
		such that the number of clusters $N_{\text{\textsf{cl.}}}=L/n$ is meaningfully defined as an integer.
		Similar conditions arise for other lattices when using non-overlapping clusters. In case of the square link lattice,
		the vertex and plaquette clusters are sufficient to learn the constraints $A_v=1$ and $B_p=1$, respectively.
		The most unbiased cluster choice when no information about the Hamiltonian is available, however,
		is a multiple of unit-cells of the lattice such as the 8-site cluster, which is large enough to capture both
		local constraints simultaneously.}
\label{fig:clusters}
\end{figure*}

For a fixed choice of hyper-parameters every Bloch-vector configuration is partitioned into clusters
$\{S^c_i\} \longmapsto \{S_{J,\alpha}^c\}$ as described in sec.~\ref{sec:Method} for the chain lattice. 
Fig.~\ref{fig:clusters} shows several possible clustering choices for the chain and square link lattice.
The feature construction used in TK-SVM makes use of the fact that local order parameters of symmetry-broken
phases~\cite{Greitemann_2019} and entanglement structures of SPT phases~\cite{Chen13} can be
represented by a finite number of elementary degrees of freedom.
The feature mapping provides a high-dimensional vector space to host potential local orders
and entanglement structures, while SVM optimizations search for their explicit expressions.
The results of the SVM optimization are weights ($\lambda_s$) for the feature vectors
$\vekk{\phi}^{(s)}$, from which the coefficient matrix $C_{\mu\nu} = \sum_s \lambda_s y_s \phi_{\mu}^{(s)} \phi_\nu^{(s)}$
(c.f. Eq.~\eqref{eq:decisionFunc_3})) is constructed.
Each row or column of $C_{\mu\nu}$ encodes pair-wise correlations between a feature $\phi_{\nu}$ and the entire set of features $\{\phi_{\mu}\}$.
Hence, to find the non-trivial features, it typically suffices to construct a few non-vanishing rows or columns of the matrix.
Without loss of generality, one can start by examining a single column with strong overall weight $\big|\sum_s \lambda_s \phi_{\nu}^{(s)}\big|$.
The number of columns required for information convergence may depend on the nature of an order but is in principle bounded by the size of the unit cell.
In case of SPT phase in the cluster-state model and the toric code (Fig.~\ref{fig:FeatureVectors} and Fig.~\ref{fig:toricFeatureVectors} in the main
text), a single non-vanishing column is enough for extracting the features.

\section{Graph Partitioning} \label{sec:GraphPartitioning}

A weighted graph $G=(V,E,w)$ is defined as a tuple of a set of vertices $V$, a 
set of edges $E$ and a set of normalized weights $w$ on the edges. All graphs
produced by TK-SVM are of a restricted class of undirected graphs with no self-edges 
(edges that connect a vertex to itself) and 
no multi-edges ($E$ is not a multiset).\\
The goal is to find a two-way partition $V_1\cup V_2=V\ |\ V_1\cap V_2=\emptyset$, 
such that $|V_1|\approx |V_2|$ and the connectivity, i.e. the summed weight of all edges connecting
$V_1$ and $V_2$ is minimal. Fiedler's theory~\cite{Fiedler73,Fiedler75} makes use of the Laplace matrix 
\begin{align}
    \Lambda_{i\neq j} = 
    \begin{cases}
        -w_{ij}, &\text{if } (i,j) \in E\\
        0,       &\text{otherwise}
    \end{cases}, \quad\quad
    \Lambda_{ii} = \sum_{\{j|(i,j)\in E\}} w_{ij}.
\end{align}
Further define a partition vector as $\vekk{x}=\{x_i=\pm1\}$ where a positive (negative) sign indicates that
the vertex $i$ belongs to $V_1$ ($V_2$).
For simplicity we assume that the graph is connected, as the discussion of multi-component graphs can always be
reduced to discussing the disconnected components separately.\\
For an arbitrary partition vector $\vekk{x}$, the quadratic form 
$\vekk{x}^T\Lambda \vekk{x}$ equals four times the connectivity
\begin{equation}
\vekk{x}^T \Lambda \vekk{x} =\sum_{(ij)\in E} w_{ij}(x_i-x_j)^2 = 4\sum_{(ij)\in E_{\text{con}}} w_{ij},
\end{equation}
where $E_{\text{con}}\subset E$ is the set of edges that connect $V_1$ and $V_2$.
So the goal is to find a partition vector that minimizes the quadratic form 
while maintaining (approximately) equally sized parts.
By changing from discrete variables $x_i\in\{-1,1\}$ to continuous ones $z_i\in[-1,1]$ and diagonalizing
$\Lambda$, it can be shown that
\begin{equation}
\min(\vekk{z}\Lambda \vekk{z}) = |V|\ \lambda_2,
\end{equation}
with $\lambda_2$ being the second smallest of all eigenvalues of $\Lambda$. Note that any Laplace matrix has 
non-negative eigenvalues. Thus the argument minimizing the quadratic form is simply the second eigenvector $\vekk{z}_{(2)}$
of $\Lambda$. To recover a discrete solution from the optimal continuous partition vector $\vekk{z}_{(2)}$, the sign function is applied 
to each component
\begin{equation}
    \vekk{x}_\text{min}=\text{sign}(\vekk{z}_{(2)}).
\end{equation}
For a general multi- rather than two-way partition of the graph, in place of the sign function applied to the Fiedler vector,
clustered appearances of similar entries can indicate more than two parts.
In applications of TK-SVM, the final number of parts of the graph is unknown, requiring a multi-way partition and hence a
continuous representation of the Fiedler vector. The lower panel of Fig.~\ref{fig:phaseDiagram} displays the histogram
of Fiedler vector entries as obtained from partitioning the graph constructed for the cluster model.\\

In the TK-SVM framework, we relate the learning of an unknown phase diagram to a graph partitioning problem. 
Specifically, each dataset sampled at some physical parameter is represented as a vertex in the graph.
Edges connecting two vertices are weighted by the bias parameter $b$, see Eq.~\eqref{eq:BiasCriterion}.
The weights are normalized using a Lorentzian function
\begin{align}\label{eq:weight}
	w(e_b) = 1-\f{b_c^2}{(|b|-1)^2 + b_c^2} \in [0,1),
\end{align}
where the parameter $b_c$ sets a characteristic scale quantifying ``$\gg 1$".
As we consider a fully connected graph, edges between datasets located deep in phases are also included.
This redundancy makes the choice of $b_c$ uncritical: the partitioning is typically robust against varying $b_c$ over
several orders of magnitude~\cite{Greitemann19b}.

\section{Benchmarking Prediction Accuracy} \label{sec:Accuracy}
\begin{figure}[t]
  \centering
  \includegraphics[width=0.5\textwidth]{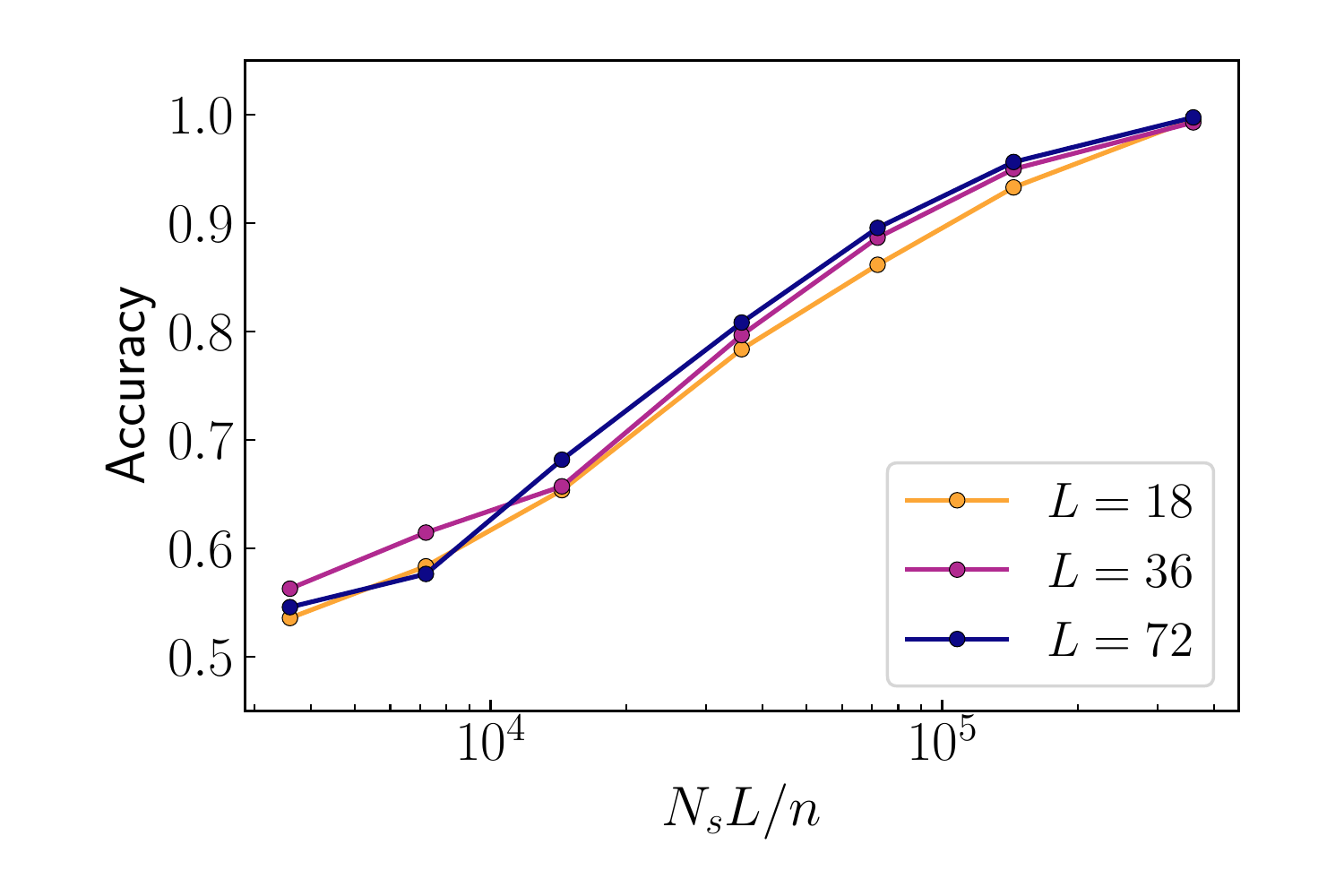}
  \caption{Dependence of binary prediction accuracy on the number of samples ($N_s$) and system size ($L$). The $x$-axis is scaled with the number of local clusters, $L/n$, in the sample. Samples are collected from the clean cluster state and classified against random samples. A $r=n=5$ TK-SVM with overlapping clusters is employed.}
  \label{fig:accuracy_lengths}
\end{figure}
\begin{figure}[t]
  \centering
  \includegraphics[width=0.5\textwidth]{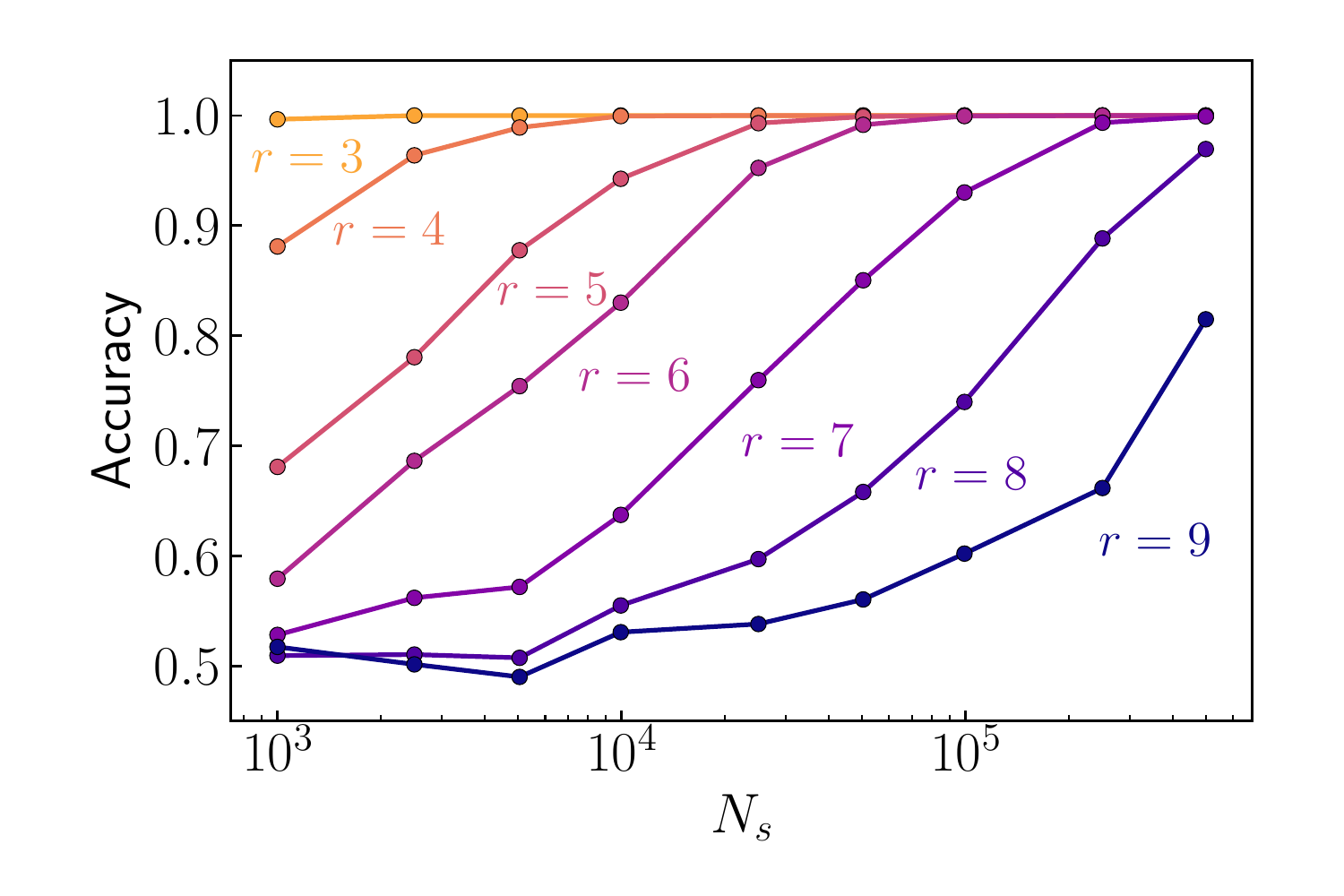}
  \caption{Dependence of binary prediction accuracy on the ranks and number of samples for a system of size $L=72$, using overlapping clusters. 
  	An accuracy $\sim 0.8$ is typically enough to learn a clear structure for the underlying order.}
  \label{fig:accuracy_ranks}
\end{figure}
Here we investigate the learning efficiency of our machine as a function of system size and number of samples.
For simplicity, we restrict ourselves to train the cluster state of length $L$ against random samples.
The learning success is determined by the accuracy with which the machine correctly classifies data from an unseen testset.
The benchmarks in this section aim to understand how many samples are sufficient for our machine to learn
the characteristic observables of an entangled quantum state.

We first discuss the dependence on system size with fixed rank $r$ and cluster size $n$.
As displayed from the coinciding data points for $L=18,36,72$ in Fig.~\ref{fig:accuracy_lengths}, our machine shows a (nearly)
constant complexity over the total number of clusters $N_sL/n$: Fewer samples are needed for larger systems to reach a certain accuracy.
This is understood from the feature construction Eq.~\eqref{eq:phiJ}: Larger systems have more local clusters to average over,
yielding smaller errors of the corresponding feature.

Next, we discuss the effects of the rank $r$, which detects $r$-body quantities.
We choose $r=n$ for simplicity and analyze the prediction accuracy up to rank $r=9$.
As shown in Fig.~\ref{fig:accuracy_ranks}, the required number of samples to reach good accuracy scales exponentially with $r$
as expected, since the dimension of the feature space is $\propto 3^r$.
This has a clear physical implication: In general, the efficiency of an unbiased machine learning algorithm strongly depends
on the nature of the phase and the complexity of the target variable.
Few samples are typically enough for learning the simplest orders, but complicated orders and entanglement patterns require more.
It also represents an intrinsic difficulty for unbiased algorithms to learn arbitrarily high-rank quantities and long-range entanglement structures.

\bibliography{qtksvm_p1}

\begin{thebibliography}{49}
\bibitem [{not()}]{note}%
  \BibitemOpen
  \href@noop {} {}\bibinfo {note} {We note that, in the machine-learning
  literature, the term ``classifying phases'' typically refers to
  discriminating the phases in given datasets instead of a topological or
  symmetry classification.}\BibitemShut {Stop}%
\end{thebibliography}

\end{document}